%%%%%%%%%%%%%%%%%%%%%%%%%%%%%%%%%%%%%%%%%%%%%%%%%%%%%%%%%%%%%%%%%%%%%%%%%%
%                                                                        %
%      Fusion rules for admissible representations of affine algebras:   %
%            the case of $A_2^{(1)}$                                     %
%                                                                        %
%         by P. Furlan, A.Ch. Ganchev and V.B. Petkova                   %
%                                                                        %
%                                                                        %
%          TEX file using harvmac.tex, amssym.def, amssym.tex;           %
%          TEX file with figures using XY-pic package.                   %
%                                                                        %
%                             hep-th/9709103                             %
%                                                                        %
%%%%%%%%%%%%%%%%%%%%%%%%%%%%%%%%%%%%%%%%%%%%%%%%%%%%%%%%%%%%%%%%%%%%%%%%%%
%

\input harvmac.tex
\hfuzz 15pt
%\draftmode

%Macros
%%%%%%%%%%%%%%%%%%%%%%%%%%%%%%%%%%%%%%%%%%%%%%%%%%%%%%%%%%%
%
\def\fwt{\omega}   
\def\hsl3{\widehat{sl}(3)} 
\def\wyg{{\cal W}} % Weyl graph
\def\fwd{{\cal G}} % finite generalised weight diagram
%
%%%%%%%%%%%%%%%%%%%%%%%%%%%%%%%%%%%%%%%%%%%%%%%%%%%%%%%%%%%%

%%%%%%%%%%%%%%%%%%%DEFINITIONS%%%%%%%%%%%%%%%%%%%%%%%%%%%%%%%%%

\def\za{\alpha} \def\zb{\beta} \def\zg{\gamma} \def\zd{\delta}
   
\def\zl{\lambda} \def\zm{\mu} \def\zn{\nu} \def\zo{\omega}
  \def\zs{\sigma}

\def\h#1{\hat{#1}}
\def\hi{{\hat\imath}} \def\hj{{\hat\jmath}}

\def\zL{\Lambda}  

\def\IZ{Z\!\!\!Z}
\def\[{\,[\!\!\![\,} \def\]{\,]\!\!\!]\,}
\def\dC{C\kern-6.5pt I}

\def\la{\langle} \def\ra{\rangle}

\def\bw{\bar w}
\def\ba{\bar \alpha}
\def\bS{\bar S}

\def\bW{\overline W}

\def\sl{sl(3)}

\def\CG{{\cal G}}

\def\N{\IN}

\def\G{\CG}

\def\un{{\bf 1}}

%xxxxxxxxxxxxxxxxxxxxxxxxxxxxxxxxxxxxx
%

%\def\pd{\partial}
%\def\der#1{{\partial \over \partial #1}}
%\def\dd#1#2{{\partial #1 \over \partial #2}}

\def\({ \left( } %\def\[{ \left[ }
\def\){ \right) } %\def\]{ \right] }

\def\mod{{\rm mod\,}}

\catcode`\@=11
\def\Eqalign#1{\null\,\vcenter{\openup\jot\m@th\ialign{
\strut\hfil$\displaystyle{##}$&$\displaystyle{{}##}$\hfil
&&\qquad\strut\hfil$\displaystyle{##}$&$\displaystyle{{}##}$
\hfil\crcr#1\crcr}}\,}   \catcode`\@=12

%\font\HUGE=cmbx12 scaled \magstep4
%\font\Huge=cmbx10 scaled \magstep4
%\font\Large=cmr12 scaled \magstep3
%\font\MLarge=cmti12 scaled \magstep3
%\font\large=cmr17 scaled \magstep0
%\font\Gros=cmbx12 scaled 1200 %

%%%%%%%%%%%%%%%%%%%%%%%%%%%%%%%%%%%%%%%%%%%%%%%%%%%%%%%%%%%%

%%%%%%%%%%%%%%%%%%%%%%% amsTEX characters %%%%%%%%%%%%%%%%%%

\def\IC{\relax\hbox{$\inbar\kern-.3em{\rm C}$}}

%\def\msy{y }
%\message{Do you have the AMS fonts (y/n) ?}\read-1 to \msan\ifx\msan\msy
%\input mssymb
\input amssym.def
\input amssym.tex
\def\IZ{\Bbb Z}\def\IC{\Bbb C}\def\IN{\Bbb N}
 
%\else
%\def\gg{g}
%\fi

% $\IC$

%%%%%%%%%%%%%%%%%%%%%%%%%%%%%%%%%%%%%%%%%%%%%%%%%%%%%%%%%%%%%%%%%

% References

% Singular vectors, adm reps

\lref\KW{V.G. Kac and M. Wakimoto,  Proc. Natl. Sci. USA {\bf 85}
         (1988) 4956  \semi 
   V.G. Kac and M. Wakimoto, Adv. Ser. Math. Phys. {\bf 7} (1989) 138
         \semi
   V.G. Kac and M. Wakimoto, Acta Applicandae Math. {\bf 21} (1990) 3.}
%\lref\FKW{ E. Frenkel, V. Kac and M. Wakimoto, Comm. Math. Phys.
%{\bf 147} (1992) 295.}

\lref\MFF{F.G. Malikov, B.L. Feigin and D.B. Fuks, Funkt. Anal.
          Prilozhen. {\bf 20}, no. 2 (1987) 25.}

\lref\KK{V.G. Kac and D.A. Kazhdan, Adv. Math. {\bf 34} (1979) 97.}

\lref\VD{V.K. Dobrev, {\it Multiplet classification of the
    indecomposable highest weight modules over affine Lie algebras
    and invariant differential operators: the $A_l^{(1)}$ example},
    ICTP preprint IC/85/9.}  

\lref\AY{H. Awata and Y. Yamada,  Mod. Phys. Lett. 	{\bf A7}
      (1992) 1185.} 

\lref\FM{ B.L. Feigin and F.G. Malikov,  Lett. Math. Phys.
	{\bf 31} (1994) 315; {\it Modular functor and representation 
	theory of $\hat{sl(2)}$ at a rational level}, q-alg/9511011.}

\lref\FGPc{ P. Furlan, A.Ch. Ganchev and V.B. Petkova, Nucl. Phys.
            {\bf B 431} (1994) 622.} 
			
\lref\FGPa{ P. Furlan, A.Ch. Ganchev and V.B. Petkova, Nucl. Phys.
            {\bf B 491} [PM] (1997) 635.} 

\lref\Zh{ D.P. Zhelobenko,  {\it Compact Lie Groups and their
        Representations}, (Amer. Math. Soc., Providence, 1973,
        translated from the Russian edition, Nauka, Moscow, 1970).}

% Verlinde formula

\lref\BPZ{A. Belavin, A. Polyakov and A. Zamolodchikov, 
          Nucl. Phys. {\bf B241} (1984) 333.}

\lref\V{E. Verlinde, Nucl. Phys. {\bf B300} [FS22] (1988) 360.}

\lref\K{V.G. Kac, {\it Infinite-dimensional Lie Algebras}, third
        edition (Cambridge University Press, Cambridge 1990.}

\lref\MW{ M. Walton, Nucl.Phys. {\bf B 340} (1990) 777.}

\lref\FGP{ P. Furlan, A.Ch. Ganchev and V.B. Petkova, Nucl. Phys.
           {\bf B 343} (1990) 205.} 

% Graphs

\lref\Pas{V. Pasquier,  J. Phys. {\bf A20} (1987) 5707.}
   %{\it Nucl. Phys.} {\bf B285} [FS19] (1987) 162 \semi V.
   %Pasquier,   

\lref\DFZ{P. Di Francesco and J.-B. Zuber,
          Nucl. Phys. {\bf B338} (1990) 602%.} 
  \semi
P. Di Francesco and J.-B. Zuber,
in {\it Recent Developments in Conformal Field Theories}, Trieste
Conference 1989, S. Randjbar-Daemi, E. Sezgin and J.-B. Zuber
eds., World Scientific 1990  \semi 
P. Di Francesco,  Int. J. Mod. Phys. {\bf A7} (1992) 407.}

\lref\BI{ E. Bannai, T. Ito, {\it Algebraic Combinatorics I: Association
Schemes}, Benjamin/Cummings (1984).}

\lref\JB{J.-B. Zuber,  Comm. Math. Phys. {\bf 179} (1996) 265. }
% \semi
%J.-B. Zuber, {\it Generalised Dynkin diagrams and root systems
% and their folding}, hep-th/9707....}

%%%%%%%%%%%%%%%%%%%%%%%%%%%%%%%%%%%%%%%%%%%%%%%%%%%%%%%%%%%%%%%%%%%%%

%\Title{%\vbox{\hbox{

\rightline{ \tt hep-th/9709103}
\vskip 2cm

\centerline{\bf Fusion rules for admissible representations of} 
\centerline{\bf affine algebras: the case of $A_2^{(1)}$} 

\bigskip\bigskip
\centerline{P. Furlan$^{*, \dagger}\, $, $\ $  A.Ch.
Ganchev$^{**, \dagger,\sharp}$ $ \ $
 and $\ $ V.B. Petkova$^{**, \dagger}$} 
\bigskip
\centerline{\it Dipartimento  di Fisica Teorica
 dell'Universit\`{a} di Trieste, Italy$^{*}$,}
\medskip
\centerline{\it Istituto Nazionale di Fisica Nucleare (INFN),
Sezione di Trieste, Italy$^{\dagger}$,}
\medskip

\centerline{\it Institute for Nuclear Research and
Nuclear Energy, Sofia, 
Bulgaria$^{**}$}
%\centerline{\it 72 Tsarigradsko Chausee, 1784 Sofia,
%Bulgaria$^{**}$} 
\medskip

\centerline{\it FB Physik, Uni Kaiserslautern, Germany$^{\sharp}$}

\bigskip\bigskip

\vskip .2in

%\noindent 
{\bf Abstract}
\medskip

We derive the fusion rules  for a basic series of admissible
representations of $\widehat{sl}(3)$ at fractional level $3/p-3$.  
The formulae admit an interpretation in terms of the affine Weyl
group introduced by Kac and Wakimoto. It replaces the ordinary
affine Weyl group in the analogous formula for the fusion rules
multiplicities of integrable representations.  Elements of the
representation theory of a hidden finite dimensional graded algebra 
behind the admissible representations are briefly discussed.

\bigskip

Pacs: 11.25.Hf; Keywords: fusion rules, admissible representations.
\bigskip

%\draftmode
%

%%%%%%%%%%%%%%%%%%%%%%%%%%%%%%%%%%%%%%%%%%%%%%%%%%%%%%%%%%%%
%                       SECTION ONE: INTRO
%%%%%%%%%%%%%%%%%%%%%%%%%%%%%%%%%%%%%%%%%%%%%%%%%%%%%%%%%%%%

\newsec{Introduction}
%
%\nind 
The fusion rules (FR)  are basic ingredient in any 2-dimensional
conformal field theory \BPZ, \V.  In \AY\ Awata and Yamada
derived the FR for admissible irreducible representations of
$\widehat{sl}(2)$ characterised by rational values of the level
and the highest weights. The aim of this work is to extend this
result to higher rank  cases -- for simplicity  we present here
the case   $\hsl3$. More details and the general case
$\widehat{sl}(n)$  will appear  elsewhere.

There is a formula for the FR multiplicities \K, \MW, \FGP\ in
the case of integrable representations of $\widehat{sl}(n)$,
equivalent to the better known Verlinde formula.  It generalises
the classical expression for the multiplicity of an irreducible
representation in the tensor product of two finite dimensional
$sl(n)$ representations, resulting from the Weyl character formula.
The  derivation in \FGP\ was based essentially on the
fundamental role played by the representation theory of $sl(n)$
and of its quantum counterpart $U_q(sl(n))$ at roots of unity.
What complicates the problem under consideration is precisely the
lack of knowledge of what is the finite dimensional algebra  and
its quantum counterpart,  whose representation theory lies behind
the fusion rules of admissible representations. In the case of
$\widehat{sl}(2)$ Feigin and Malikov \FM\ have noticed that the
relevant  algebra is the superalgebra $osp(1|2)$ and its deformed 
counterpart. Inverting somewhat the argumentation in
\FGP,  we shall try to show that the understanding of the fusion
rules for admissible representations leads naturally to a set of
finite dimensional representations of some (graded) algebra with
well defined ordinary tensor product.

%%%%%%%%%%%%%%%%%%%%%%%%%%%%%%%%%%%%%%%%%%%%%%%%%%%%%%%%%%%%
%                       section 2: 
%%%%%%%%%%%%%%%%%%%%%%%%%%%%%%%%%%%%%%%%%%%%%%%%%%%%%%%%%%%%
\newsec{Admissible weights.}

We start with introducing some notation.
The simple roots of $\hsl3$ are $\za_i$, $i=0,1,2\,$. The affine
Weyl group $W$ is generated by the three simple reflections
$w_i=w_{\za_i}$. For short let $w_{ij\dots}=w_i w_j \dots$ and
furthermore $w_{\h1}=w_{020}=w_{202}$, $w_{\h2}=w_{010}=w_{101}$,
$w_{\h0}=w_{121}=w_{212}\,$, corresponding to the real positive
roots $\za_{\hi}=\delta -\za_i\,,$
$\delta=\za_0+\za_3=\za_0+\za_1+\za_2\,.$ The elements $w_{\hi
i}$ and $w_{i \hi}$ constitute (elementary) affine translations
$t_{\pm \ba_i}\,,$ where   $\ba_0=-\za_3\,,
\ba_i=\za_i\,, i=1,2\,.$
Denote the fundamental weights  by $\fwt_0\,,\,
\fwt_i+\fwt_0\,$, $\,i=1,2\,$, $\fwt_i$ being the horizontal
$\sl$ subalgebra fundamental weights.  Let $P=\sum_{i=1}^2
\IZ\,\fwt_i\,,$ $P_+=\sum_{i=1}^2 \IZ_+\,\fwt_i\,,$
$P_{++}=\sum_{i=1}^2 \IN\,\fwt_i$ be the  $\sl$ weight lattice,
the integral dominant  and strictly integral dominant  weights
and  $P^k_+=\{ \zl\in P_+\,, \, \la \zl,
\za_3 \ra \le k  \}\,$, $P^{k+3}_{++}=\{ \zl\in P_{++}\,, \, \la
\zl, \za_3 \ra \le k+2  \}\,$  ($k\in \IZ_+)\,,$  the
corresponding alcoves. For short we shall denote all $\hsl3$
weights at an arbitrary level $k$  with their horizontal
projections $\zL=\sum_{i=1,2}\, \la \zL,\za_i\ra\, \fwt_i\,.$
Accordingly $w\cdot \zL$ will indicate the horizontal projection
of the shifted (by $\,\fwt_1+\fwt_2+3\,\fwt_0$) action on 
$\zL+ k \omega_0\,$ of the
affine Weyl group. For nonnegative integer $k\,$ $\,\zl
\in P^k_{+}$ ($P^{k+3}_{++}$) runs over highest weights (shifted
by $\rho=\fwt_1+\fwt_2$ highest weights)  of integrable
representations at level $k$.

Given a fractional level $k$ such that $\kappa\equiv k+3 =p'/p$
with $p,p'$ coprime integers and $p' \ge 3\,,$ the set of 
admissible weights of $\hsl3$ is defined \KW\ as\foot{We neglect 
the terms $d \delta$ in the full admissible highest weights as 
irrelevant to our purposes.}
%the (disjoint) union
%
\eqn\da{
P_{p',p} = \{ \zl' - \zl \, \kappa\, | \, \zl'\in
              P_+^{p'-3}\,, \ \zl \in  P_+^{p-1}\}   
     \cup  \{
w_3\cdot (\zl' - \zl \, \kappa)\, |\, \zl'\in
              P_+^{p'-3}\,, \  \zl \in  P_{++}^{p+1}\}\,.
}
Due to invariance with respect to a Coxeter element
generated subgroup of the horizontal Weyl group
$\bW$ (see \KW\ for details) the domain in \da\ 
can be equivalently represented using other elements $w\in \bW$.

We shall refer to $P_{p',p}$ as the admissible alcove and to its
first, second subsets as its first, second leaf. We have reversed 
the traditional notation putting the prime on the integer part of 
the  weights and leaving the fractional part unprimed,  the reason 
being that in this paper we shall restrict ourselves mostly to the 
particular series of admissible representations defined by $p'=3$, 
$p\ge 4\,,$ in which only  $\lambda'=0$ survives in the pairs
$(\zl',\zl)$ of $sl(3)$ weights appearing in \da. The
addmissible alcove $P_{3,p}$  (to be called sometimes ``the
double alcove'') is described by a collection of ${p+1\choose 2}+
{p\choose 2}=p^2\,$ integrable weights at integer levels $p-1$
and $p-2$, entering the fractional parts of the weights of the
first and second leaf respectively. The choice $p'=3$ is not very
restrictive since the novel features of the fusion rules are
essentially governed by the fractional part of the admissible
highest weights and furthermore the subseries $p'=3$ is
interesting by itself. Shorthand notation $[n_1,n_2]$ and $\[
n_1,n_2 \] \,$, $n_i=\la\zl, \za_i \ra \,$, will be used for the
weights $\zL=-\zl \kappa\,$ on the $1^{\rm st}$ leaf and
$\zL=w_3\cdot(-\zl \kappa)\,$ on the $2^{\rm nd}$  leaf of
$P_{3,p}\,$ in \da,  respectively; $n_3=n_1+n_2$. For $\zL\in
P_{3,p}$ we shall exploit the automorphism groups $\IZ_3$ of the
alcoves $P^{p-1}_+$ and $P^{p+1}_{++}$ generated by 
\eqn\aut{\eqalign{
\zs:\ \zL&=[n_1, n_2] \mapsto \zs(\zL) =
[p-1-n_3, n_1]\,, \cr
\zs:\ \zL&=\[n_1, n_2\] \mapsto \zs(\zL) =
\[n_2, p+1-n_3 \]\,. 
}}

The $\hsl3$ Verma modules labelled by  admissible highest weights 
are reducible, with submodules determined by the Kac-Kazhdan 
theorem \KK.  In general for an arbitrary weight $\zL$ let $M_i=
\la \zL +\rho + \kappa \fwt_0, \za_i\ra\,$, $i=0,1,2\,$.  If for 
a fixed $i=0,1,2\,,$ the  projection  $M_i$ can be written as
$M_i=m'_i-(m_i-1)\,\kappa\,$ (or as $M_i=-m'_i+m_i\,\kappa\,$),
where $m'_i\,, m_i \in \N \,$,  there is a singular vector  of
weight $w_{\zb_i} \cdot \zL$ (or $w_{\zb_\hi} \cdot \zL\,$) in
the Verma module of highest weight $\Lambda$. It corresponds to
the affine real positive root $\zb_i=(m_i-1)\,
\delta+\za_i\,$ (or $\zb_\hi=m_i\, \zd -  \za_i\,$) respectively. 
These Weyl reflections can be represented as 
\eqn\we{
w_{\zb_i}
=t_{-(m_i-1)\,\ba_i}\,w_{i}
=w_i(w_{\hi i}
)^{m_i-1}\,, \quad
 w_{\zb_{\hi}}
=t_{(m_i-1)\,\ba_i}\, w_{\hi}
=w_{\hi} 
(
w_{ i\hi })^{m_i-1}\,, \quad i=0,1,2\,.
}
The corresponding  singular vectors  were constructed in \MFF. 
To a decomposition of the reflections \we\ into a product of 
simple reflections  corresponds a monomial of the lowering
generators of $\hsl3$, namely every $w_i$, $i=0,1,2$, is 
substituted by $E^{-\za_i}$ to an appropriate (in general 
complex) power, see \FGPc\ for more explicit presentation 
in the case of $\hsl3$.

Consider the Weyl groups $\bW^{\zl}$ and $W^{\zl}$ generated for
the representations on the first  (second) leaf of \da\ by the
%simple 
reflections $w_{\zb_i} (w_{\zb_\hi})\,, i=1,2\,,$
$\zb_i=\la\zl,\za_i \ra\, \zd+\za_i\,$
($\zb_\hi=-\la\zl,w_3(\za_i) \ra\,  \zd -\za_i\,$ ) and
$w_{\zb_i}$ ($w_{\zb_\hi})\,, i=0,1,2\,,$ with $\zb_0
=(p-1-\la\zl,\za_3 \ra ) \delta +\za_0$ ($\zb_{\h0}=(p+1-\la\zl,\za_3
\ra ) \delta -\za_0$) respectively. Similarly one defines 4 more
variants corresponding to the various equivalent ways of
representing the admissible alcove.  The groups $\bW^{\zl}$ and
$W^{\zl}$  are isomorphic to $\bW$ and $W$ and were introduced
and exploited by Kac and Wakimoto (KW) \KW\ in the study of
the characters of admissible representations.  We shall refer to
these groups, which will play a crucial role in what follows, as
the KW groups.  Apparently any element of the KW group $W^{\zl}$
depends on (and is determined by) the point on which it acts, so
in a sense this is a ``local'' group acting on the double alcove
and ``spreading'' it much in the same way as the ordinary affine
Weyl group acts on the fundamental integrable alcove. This is
more transparent using an alternative description of  $P_{3,p}$.
 
\medskip

%%%%%%%%%%%%%%%%%%%%%%%%%%%%%%%%%%%%%%%%%%%%%%%%%%%%%%%%%%%%
%                       section 3: 
%%%%%%%%%%%%%%%%%%%%%%%%%%%%%%%%%%%%%%%%%%%%%%%%%%%%%%%%%%%%

\newsec{ Alternative description of the admissible alcove
$P_{3,p}\,.$ 
Affine Weyl group graph replacing the weight lattice.}

First recall that the affine Weyl group $W$ can be represented as
a graph to be denoted $\wyg$. This is the well known ``honeycomb''
lattice (which we saw for the first time in \VD$\,$) covering 
the plane, consisting of hexagons with links labelled by the
three elementary reflections $w_i$, $i=0,1,2$, -- so that at each
vertex meet 3 (always different) links, while along any hexagon
two of the three reflections appear at an alternating order.  
(All figures in this paper depict some finite part of this graph;
the labels 0,1,2 on the edges correspond to the reflections
$w_0,w_1,w_2$.) The three different hexagons appearing on the
graph depict the three 6-term relations among the generators
$w_i$, $i=0,1,2$, e.g., $w_{121}=w_{212}$. Thus if we choose an
origin, a vertex of this graph, to represent the identity element
of $W$, then the vertices are in bijective correspondence with
the elements of $W\,,$ while the different paths connecting a
vertex with the origin depict the different presentations of an
element of $W$ in terms of the generators $w_i$.

The dual lattice has as vertices the centers of the hexagons, 6
dual links meet at a dual vertex, and the elementary cells are
triangles centered at the vertices of the hexagonal lattice.  For
given $\kappa=3/p$ consider a triangle in the dual lattice the
sides of which consist of $p$ dual links. The 
weights in $P_{3,p}$ can be arranged on the piece of $\wyg$
 enclosed by this triangle.  This ``big'' alcove consists of
$p^2$ elementary triangles, equivalently $p^2$ vertices of the
honeycomb. The three vertices sitting at the three corners are
connected with the other points of the  alcove
by the three possible links $0,1,2$. Choose the corner vertex
that is connected by a 0-link as the origin and assign to it the
weight [0,0]. To the other vertices of the alcove assign weights
by the shifted action of $W$. The vertices at an even number of
links from the origin accommodate weights on the $1^{\rm st}$ leaf
while the ones at an odd number -- on the $2^{\rm nd}$ leaf. The
example $p=4$ is depicted on fig. 1.  The  three lines of the dual 
lattice cutting out the alcove are indicated by dotted lines.  
The weights $[n,m]$ are in 
circles, the weights $[[n,m]]$ in boxes. (In general the first few 
points are e.g.,  
$w_{20}\cdot [0,0]=[p-2,1]\,,$  $w_{120}\cdot [0,0]=\[1,p-2\]\,,$ 
$w_{0120}\cdot [0,0]=[0,p-3]\,,$ etc., obtained by
adding and subtracting $p'=p \kappa$.) 
The border lines (paths) of the big alcove consist of 
highest weights $\zL$ such that $w_i\cdot \zL
\not \in P_{3,p}\,,$ or $w_{\hi}\cdot \zL \not \in P_{3,p}\,,$
for some $i=0,1,2$ -- they precisely exhaust the weights
labelling the border points of the two alcoves $P^{p-1}_+$,
$P^{p+1}_{++}$. 

``Reflecting'' $\Lambda$ in the three boundary lines enclosing 
the alcove we land at a weight of a singular vector in the 
$\widehat{sl}(3)$ Verma module of highest weight $\Lambda$. These 
reflections generate a group that  coincides with $W^{\zl}$. The 
alcove is a fundamental domain with respect to the action of this 
group. In this realisation the affine Weyl group graph
$\wyg$ plays the role of  ``weight lattice''. The choice of the
reflecting ``hyperplanes'' described by $\la \zL'+\rho +\kappa \zo_0,
\zb_i \ra =0$ and $\la \zL'+\rho+\kappa \zo_0, \zb_{\hi} \ra =0\,,$
 $i=0,1,2$
(i.e., their precise identification with the three planes cutting
the big triangle) depends on the value of $p$ and the given
highest weight $\zL$ in the alcove.

As it is clear from fig. 1. any ``big'' alcove $P_{3,p}$ can be
canonically mapped into $P_{++}^{3 p}\,,$ the admissible highest
weights being identified with a subset of the triality $0$
integrable (shifted by $\rho$) highest weights at level $3p-3$.

\medskip

%%%%%%%%%%%%%%%%%%%%%%%%%%%%%%%%%%%%%%%%%%%%%%%%%%%%%%%%%%%%
%                       section 4: 
%%%%%%%%%%%%%%%%%%%%%%%%%%%%%%%%%%%%%%%%%%%%%%%%%%%%%%%%%%%%

\newsec{Decoupling of states generated by singular vectors.
Affine Weyl group graph replacing the root lattice.}

In view of the complexity of the Malikov-Feigin-Fuks (MFF)
singular vectors in the case $\widehat{sl}(n)\,$, $n>2\,$, it is
not realistic to repeat the derivation for $n=2$ in \AY, by
solving all equations for the 3-point blocks expressing the
decoupling of the submodules generated by  singular vectors in
the admissible Verma modules.  Instead we shall reduce the
problem to the solution of some basic subset of equations.

Our first step is the solution of the simplest equations arising
from the singular vectors in the Verma module labelled by the
``fractional fundamental'' weight $f:=[1,0]=-\kappa\fwt_1$ (for
short we will call it fundamental).  Using the standard
realisation of the induced representations of $SL(3)$ in terms of
functions $\varphi(x,y,z)$ depending on (``isospin'') coordinates
$(x,y,z)$, described by a triangular matrix in the  Gauss
decomposition of the elements of $SL(3)$,  we realise the
generators by the corresponding differential operators (see,
e.g., \Zh).  The isospin  coordinates (as well as the space-time
coordinates) of the  first  and third fields in the 3-point
functions  are fixed to $0$ and ``$\infty$''. For the weight $f$
(labelling the field at the first point) there are two  singular
vectors corresponding to the roots $\zb_1=\delta+\za_1$ and
$\zb_2=\za_2$ and determined by \we\ with
$w_{\zb_1}=w_{12021}=w_{1 \h1 1}\,$, $w_{\zb_2}=w_{2}\,$.  Since
the second of the singular vectors reduces simply to the $sl(3)$
generator $E^{-\za_2}=-(\partial_y + x\,\partial_z) $ of
$y$-translations, which annihilates the monomials of type $x^a\,
(xy-z)^c\,,$ we look for a solution of the equation corresponding
to the first MFF operator (see \FGPc\ for details) in terms of
such monomials (in general  -- $x^a\, y^b\, (xy-z)^c\, z^d$) --
it reduces to an algebraic system of two equations for the
unknowns $(a,c)$.  Due to Ward identities corresponding to the
two Cartan generators the powers $a,b$ (or $(a,b,c,d) $ in
general) are expressed in  terms of the three weights
$\zL^{(i)}$, $i=1,2,3\,,$ labelling the fields in a 3-point
function
\eqn\cg{\eqalign{
- 2a + b - c - d + 
 \langle\zL^{(1)} +\zL^{(2)},\za_1\rangle
 & = \langle\zL^{(3)},\za_1\rangle\,, \cr
a - 2b -c - d +
 \langle\zL^{(1)} +\zL^{(2)},\za_2\rangle
 & = \langle\zL^{(3)},\za_2\rangle\,. \cr
}}

Taking $\zL^{(1)}=f$ and $\zL^{(2)}=\zL'$,
$\langle\zL^{(2)}+\rho,\za_i\rangle = M_i\,,$ $i=1,2\,,$ we find
7 solutions 
for the resulting representation $\zL^{(3)}$ in the fusion $ f
\otimes \zL'\,,$  described by the values $(a,c)=(0,0)\,$,
$(M_1-\kappa, 0)\,$, $(-M_2, M_3-\kappa)\,$, $ (0,
M_3-\kappa)\,,\, (-M_2, M_2-\kappa)\,,\, (-\kappa, 0)\,,
(0,-\kappa)\,,$ while the corresponding highest weights read:
\medskip\noindent
{\bf Proposition 1:}
\eqn\ff{\eqalign{
f \otimes \zL'&= 
(f+\zL')\oplus w_1\cdot (f+\zL')\oplus w_{21}\cdot (f+\zL')
\oplus  
w_{121}\cdot (f+\zL') \cr
&\oplus  w_{021}\cdot (f+\zL')\oplus
w_{2021}\cdot (f+\zL')\oplus w_{0121}\cdot (f+\zL')\,.
\cr}}
The  result \ff\  holds for arbitrary generic admissible highest
weight $\zL'\in P_{p',p}\,$ (i.e., such that all representations
in the r.h.s of \ff\ belong to the admisible domain \da\ ) and
arbitrary $p'$ and $p \ge 4$, i.e., we do not restrict here to
the subseries $p'=3$. Analogously the fusion rule of the
conjugate representation $f^*=[0,1]$ with a generic $\zL'\,$
reads as in \ff\ with $w_1$
and $w_2$ interchanged.

Taking $\zL'=0$, the set of $7$ weights in the r.h.s. of
\ff\ replaces the three weights of the $sl(3)$ fundamental
representation.  The latter have clear counterparts in \ff\
represented by the highest weight  $f= -\kappa\fwt_1$ and the two
last weights $w_{\h1 1}\cdot f\, =-\kappa(\fwt_1-\za_1)$, $ w_{0
\h0}\cdot f = -\kappa(\fwt_1-\za_3)\,.$\foot{The solutions of the 
equations corresponding to the singular vectors expressed by 
$E^{-\za_2}\,,\,  (E^{-\za_1})^2\,$ 
 in the $sl(3)$ case are accordingly given by the 
same monomials with $(a,c)=(0,0)\,, (1,0)\,,(0,1)\,.$ 
For a general computation of this kind, recovering the  
$sl(3)$ representation spaces, see \Zh.}

We can visualize this 7 point ``weight diagram'', to be denoted
$\fwd_f$ (or its ``shifted'' by $\zL'$  version defined by the
weights in the r.h.s. of \ff, to be denoted $\fwd_f \circ \zL'$)
as a collection of points on the affine Weyl group graph $\wyg\,,$
see fig. 2.
Identifying a reference point on this graph with a highest weight
$\zL=-\kappa \zl\,$ on the first leaf of $P_{p',p}$ 
(in our case $\zL=f$) we build its ``positive''
part ${\wyg}_{\zL}^+$ in close analogy with the set
$Q^+_{\zl}=\zl -Q^+\,,$ $Q^+=\sum_{i=1,2}
\IZ_+ \,\za_i \subset Q\,,$ $Q$ being the root lattice of
$sl(3)\,$ (see also sect. 6 below). On fig. 2 we have 
depicted the beginning of
$\wyg^+_f$. Weights from the sublattice $\Lambda - (-\kappa) Q^+$
of ${\wyg}_{\zL}^+$  are denoted by big circles. 
They correspond to sequences of 
translations  $\kappa \za_j\,, j=1,2,3\,,$ of $\Lambda$ along the
$sl(3)$ positive roots, each elementary translation
being represented by a $4$-pieces path on
${\wyg}_{\zL}^+$. Two such paths (or one on the border lines) 
replace a link on $Q^+_{\zl}$ joining two weights which differ by
a root $ -\za_j\,.$ 
Now $\fwd_f$ is the finite part of $\wyg^+_f$
consisting of all points ``enclosed'' by the three points
on $-\kappa \Gamma_{(1,0)}$, where  $\Gamma_{(1,0)}$ is the 
weight diagram of the
$sl(3)$ counterpart of $f$.  The points of $\fwd_f$ on fig. 2 are
connected  by bold links. 

On fig. 3 we have depicted the weight diagram $\fwd_{[2,0]}$. The
same picture with 
the reference point $\zL$  substituted by $\zL+\zL'$ represents
the shifted weight diagram $\fwd_{[2,0]}\circ\zL'$ obtained
solving the decoupling equations for the fusion of $[2,0]\, $
with a generic $\zL'\in P_{p',p}$.

The working hypothesis which emerges after analysing the
decoupling equations for a couple of examples  is that in general
these generalised weight diagrams $\fwd_{\zL}$ for highest weight
$\zL= -\kappa\zl$ from the $1^{\rm st}$ leaf of the admissible alcove 
are obtained according to the following rule.  Let $\Gamma_\zl\subset
Q^+_{\zl}\,$ be the  weight diagram of the $sl(3)$ irreducible 
representation of highest weight $\zl\,.$  Embed
$-\kappa\Gamma_\zl$ (as a set of points)  into the sublattice
$\zL+\kappa Q^+$ of $\wyg^+_\zL$.  Draw all paths on $
\wyg^+_\zL$, starting from the highest weight $\zL$, that lie
within the borders of $-\kappa\Gamma_\zl\,,$ including the
``border'' path connecting all points on the outmost
(multiplicity one) layer of $-\kappa\Gamma_\zl$. The diagram
$\fwd_\zL$ is the resulting finite set of  weights on $
\wyg^+_\zL$.

Besides the series $[0,n]$, or $[n,0]$, little can be said at
this point about the precise values of the multiplicities
assigned to the points of the general weight diagrams.  The
ansatz with monomials (images of $3$-point functions at $0$,
$(x,y,z)$ and ``$\infty$'') used in the analysis of the singular
vectors equations does not reproduce all expected multiplicities
and furthermore gives controversial results using different
``shifting'' representations.\foot{ On the other hand an
alternative approach based on the full invariant $3$-point
functions and singular vectors realised by ``right action''
(Verma module generating) $\widehat{sl}(3)$ generators is technically
difficult to be implemented even in the simplest examples.} The
first example  with nontrivial multiplicities is provided by
$\zL=[1,1]$ (the analog of the $sl(3)$ adjoint representation
(1,1)). Its weight diagram is depicted on {fig.} 4. To find the
indicated weight multiplicities we have followed a different
strategy which recovers as well the general FR multiplicities.

%%%%%%%%%%%%%%%%%%%%%%%%%%%%%%%%%%%%%%%%%%%%%%%%%%%
%           section 5
%%%%%%%%%%%%%%%%%%%%%%%%%%%%%%%%%%%%%%%%%%%%%%%%%%%

\newsec{Fundamental fusion graphs.
Pasquier--Verlinde type formula for the FR multiplicities.}

The rule \ff\ describes the fusion of the fundamental
representation with a generic representation $\zL'$ -- in general
one expects truncations, i.e., some of the resulting
representations in the fusion lie outside the admissible alcove
and furthermore some admissible representations are forbidden.
To determine them one has to analyse the additional
equations resulting from the decoupling of singular vectors in
the  Verma module of highest weight $\zL'$,
 taking now the fundamental one as the
``shifting'' representation.  This implies that we select and
check only the solutions of these equations corresponding to
weights belonging to the $7$ point ``shifted'' fundamental diagram
$\fwd_f \circ \zL'$; they are described by some sets of the
powers $(a,b,c,d)$ (one of them always equal to zero), subject of
the relations \cg.

Before starting this analysis we first find that the representations
 described by the $\IZ_3$
orbit of the identity element $\un=[0,0]$, i.e., the corner
points of the big alcove, are simple currents, i.e., their fusion
with an arbitrary representation on the first or second leaf of
the double alcove produces only one representation living on the
same leaf, namely the nontrivial FR multiplicities $N_{a
\sigma(\un)}^b \, $ read
\eqn\sc{
  N_{a \sigma(\un)}^b =\delta_{b \sigma (a)}\, \ \Rightarrow \ 
  N_{a \sigma(b)}^{\sigma(c)} =N_{a b}^c\,,}
where we have denoted by the same letter the action of the two
(different) $\IZ_3$ groups defined in \aut. The second equality
in \sc\ is a consequence of the first and the associativity of
the fusion rules and enables one, as in the integrable case, to
reduce the computation of the FR on any $\IZ_3$ orbit to the FR
of one representative of that orbit. The simple current property  
is established solving the relevant singular vector equations. 
E.g., for $\sigma(\un)=[p-1,0]$ they correspond to the roots 
$\zb_{\hat 1}\,, \zb_0\,, \zb_2\,$ and the affine Weyl
group elements $w_{\h1}\,, w_0\,, w_2\,,$ respectively.  

We look next at  the fusion of the fundamental representation
with the representations on the border line $[0,n]\,, 0<n<p-1$.
These representations are counterparts of the conjugates of the
symmetric representations and the intersection of their assumed
shifted weight diagrams $\fwd_{[0,n]}\circ f$ 
with the fundamental 7 point diagram $\fwd_{f} \circ [0,n]$ (both
defined by the ``shifted'' highest weight $f+[0,n]=[1,n]\,$)
contains at most 3 points corresponding to the elements $\un\,,
w_{121}\,,$ and $w_{0121}\,$ acting on the weight $[1,n]$, see
e.g. for $n=2$ the mirror counterpart of the diagram on fig. 2.
The fusion of the representations on the other two $1^{\rm st}$ 
leaf border lines ($[n_1,n_2]\,, n_2=0\,,$ or $ n_1+n_2=p-1\,$) 
are recovered by the $\IZ_3$ symmetry using $\sc$.

We next turn to the fusion of the fundamental representation $f$
with the  representations on the second leaf starting with the
point $\[1,1\]$. The two systems of algebraic equations 
corresponding to the
roots $\zb_\hi=\zd-\za_i\,,$  $i=1,2\,,$ exclude all but 3 of the
representations in \ff\ expressed by the action of $w_{21}\,, \,
w_{121}\,,$ and $w_{0121}\,$. Similarly for representations on
the border line $\[n,1\]\,$, $1<n<p-1\,$, of the second alcove 
the solution of the additional system of equations corresponding to 
the root $\zb_{\hat 1}=\zd-\za_1$ excludes the first two of the 
seven terms in \ff. Once again we extend this result to the rest
$2^{\rm nd}$ leaf border lines using the $\IZ_3$ symmetry.

The truncations described appear only on the borders of the 
double alcove (invariant under the symmetry \aut). 
Note that all representations in the r.h.s. of \ff\ belong to 
the admissible domain whenever the weight $\zL'$ in the l.h.s 
is an interior point (i.e., the interior points are generic 
points) -- this is easier
to visualize on the ``big'' alcove, see fig 1,  whose ``border''
lines accommodate all border points of the double alcove.
-- simply attach the $7$ point diagram to the given highest weight.

In principle one should look further at the solutions of
the remaining equations resulting from  more complicated MFF vectors
and check whether they lead to further truncations. 
E.g., in the case $\[n,1\]\,$ we should consider also the singular 
vector corresponding to the root $\zb_{\h2}=n \zd-\za_2\,;$
apparently the (reduced to polynomials) form of this MFF singular
 vector gets increasingly complicated for large $n$.
On the other hand it is natural to expect  that if there are 
additional truncations they  will appear for smaller $p$. So we 
have considered more thoroughly the smallest case $p=4$ where the 
MFF vectors are the simplest possible -- the outcome is that there 
are no further truncations.

The result of this analysis is summarised in the following

\medskip\noindent
{\bf Proposition 2:} \medskip\noindent
For interior points on $P_{3,p}$ the FR read
\eqn\ffa{\eqalign{
f\otimes [n_1,n_2]&=[n_1+1,n_2]\oplus [n_1-1,n_2+1]\oplus
                    [n_1,n_2-1]\oplus [n_2,p-1-n_3]\cr
                  &\oplus \[n_1+1,n_2\]\oplus \[p-1-n_3,n_1+1\]
				    \oplus \[n_2+1,p-n_3\]\,;\cr
f\otimes \[n_1,n_2\]&=\[n_1+1,n_2\]\oplus \[n_1-1,n_2+1\]\oplus 
                      \[n_1,n_2-1\]\oplus \[p+1-n_3,n_1\]\cr
                    &\oplus [n_1,n_2-1]\oplus [p-n_3,n_1-1]
					\oplus [n_2-1,p+1-n_3]\,; 
}}
On the border lines of the two alcoves constituting $P_{3,p}$ 
the FR  multiplicities read
\eqn\ft{\eqalign{
  f\otimes [0,n]&= [1,n]\oplus [0,n-1] \oplus \[1,n\]\,, 
          \ \ 0<\! n \!< p\!-\!1\,; 
          \qquad  f\otimes [0,0] = f\,; \cr
  f\otimes \[n,1\]&= \[n\!+\!1,1\]\oplus  \[n\!-\!1,2\]\oplus
          \[p\!-\!n,n\]\oplus [n,0] \oplus 
	[p\!-\!n\!-\!1,n\!-\!1] \,, \ \ 1<\! n\! <p\!-\!1\,,\cr
  f\otimes \[1,1\]&= \[2,1\] \oplus \[p-1,1\] \oplus  [1,0]\,; 
}}
the rest being determined by the $\IZ_3$ symmetries \sc.
\medskip

Following the approach of \Pas, \DFZ\ we can look at the set of
rules \ffa, \ft\ as defining for each $p$ the adjacency matrix
$G_{ab}=N_{f a}^b$ of a fusion graph, the vertices of which
correspond to 
the points of the corresponding double alcove $P_{3,p}$; we do
not indicate explicitly the links of this graph.  The complex
conjugated representation $f^*$ provides the conjugated
adjacency matrix $G^*_{ab}=N_{f^* a}^b$ describing the same graph
with inverted orientation of the links; $G^*_{a b}=G_{b a}\,.$  
Unlike the known
fusion graphs for integrable representations of $\widehat{sl}_3$
and their nondiagonal generalisations studied in \DFZ\ (see also
\JB) the graphs described by \ffa, \ft\ are not $3$--colourable
 -- if we assign the usual triality $\tau(a)=n_1-n_2$ to the
vertices $a=[n_1,n_2]$ and $a=\[ n_1,n_2 \]\, $ we will find
nonvanishing matrix elements $G_{ab}$ for some $a,b\,,$ such that
$\tau(b)-\tau(a)\not = 1$ $\mod 3$.  Nevertheless these graphs
share some of the basic features of their  integrable
counterparts, in particular the $\IZ_3$ symmetry, replaced here
by the pair \aut\ discussed above.  The matrix $G$ is normal, 
$[G,G^t]=0\,,$ and hence diagonalisable. Its
normalised eigenvectors $\psi_a^{(\mu)}\,$ 
\eqn\norm{
\, \sum_{a} \  \psi^{(\zm)}_a\,\psi^{(\zn)\, *}_a
= \zd_{\zm \zn}\,, \qquad
\, \sum_{\zm  }\  \psi^{(\zm)}_a\, \psi^{(\zm)\, *}_c =
\zd_{a c}\,,
}
labelled by some set $\{\mu \}\, $ of $p^2$ indices  can
be used to write down a  formula for the general FR multiplicities
$N_{a b}^c$ of the type
first considered by Pasquier \Pas\foot{More precisely a formula
analogous to the dual version of (5.5),  obtained by a summation
over the vertices, appeared in \Pas, producing in general 
noninteger structure constants $M_{\mu \zl}^{\nu}\,.$ }
\eqn\pvf{
N_{ab}^{\,\, c}= \sum_{\zm } \ 
{\psi^{(\zm)}_a\psi^{(\zm)}_b\psi^{(\zm)\,
*}_c \over \psi^{(\zm)}_{\un}} \,.
}
The formula can be  looked at as a
generalisation of the Verlinde formula  with the matrix 
$\psi_a^{(\mu)}$ replacing the
(symmetric) modular matrix.
In \pvf\  the vector $\psi_{\un}^{(\mu)}\,,$ the (dual)
Perron-Frobenius vector, is required to have nonvanishing entries.
In the simplest example of $p=4$ we have checked that there
exists a choice of the eigenvector matrix $\psi_{a}^{(\mu)}$
consistent with this requirement (and furthermore
$\psi_{\un}^{(\mu)} > 0\,$) and such that
\pvf\ produces nonnegative integers for the  matrix elements of
all $N_a\,$ matrices, $\,(N_a)_{b}^c=N_{a b}^c\, .$
The arbitrariness in determining the eigenvectors is caused by 
the presence of eigenvalues of $G$ of multiplicity greater than
one.
The eigenvalues $\chi_a(\zm)=\psi_{a}^{(\mu)}/\psi_{\un}^{(\mu)}$
provide $1$ - dimensional representations (``$q$-characters'') of
the algebra of $N_a$-matrices. In the $sl(2)$ case  the analogous 
to \pvf\ formula (which is more explicit since there 
is a general expression for the matrices $\psi_{a}^{(\mu)}$)
reproduces indeed the known \AY\ FR multiplicities
at level $2/p-2$.
 It should be stressed 
that the $1^{\rm st}$ leaf representation $f$ and its adjoint $f^*$ do 
not exhaust the fundamental set of fields needed to generate a fusion 
ring. However the  description of that set is not needed in 
formula \pvf, the latter providing the full  information about 
the fusion rule.  In writing \pvf\ we have essentially assumed that the 
admissible reps fusion algebra is a ``C -- algebra'' (from ``Characters 
-- algebra''),
 following the terminology in \BI,  i.e., an associative,
commutative algebra  over $\IC$ with real structure constants, 
with a finite basis, an identity element,
an involution (here the map of a weight to its adjoint) requiring some 
standard properties of the structure constants. The knowledge of the 
fundamental matrix $N_f$ specifies the algebra and allows to 
describe  the common to all
matrices $N_a$ eigenvector matrix $\psi_{a}^{(\mu)}$. 
The fact that the general formula \pvf\
for the ``C -- algebra'' structure constants gives
 nonnegative integers 
is highly nontrivial.   Comparing with  \DFZ\  (where such 
``C -- algebras'' were discussed and used), recall that
the nonnegativity of the integers in the l.h.s. of formulae 
analogous to \pvf\  selects a subclass of the graphs  related to  
 modular invariants of the integrable  models; a  counterexample, 
where this nonnegativity cannot be achieved, is provided e.g., 
by the $E_7$ Dynkin diagram.

Thus the knowledge of the fundamental fusion graph allows to 
determine in principle all FR multiplicities.

\medskip
{\bf Remark: } 
The graphs appearing in the study of the admissible
representations and the structures they determine deserve further
investigation.  In particular it is yet unclear whether  the
``dual'' version of \pvf, describing the structure constants of a 
dual ``C -- algebra'', has any importance.
It would be also interesting to check whether the set $\{\mu
\}\,, $ can be interpreted as some ``exponents'' set in the sense
of \JB.  These questions have sense already for the case of
admissible representations of $\widehat{sl}(2)$ at level $2/p-2$
where the corresponding graphs (or their unfolded colourable
ladder type counterparts) look rather simple.

\medskip
The basic FR \ffa, \ft\ which we have derived are checked to
admit another interpretation which generalises the corresponding
fusion formula of \K,\MW,\FGP\ in the case of integrable
representations.  First a representation $\zL''$ appears in the
fusion $f\otimes \zL'$ only if it  belongs to the intersection of
the $7$ points  ``shifted'' weight diagram of the fundamental
representation $\fwd_f\circ \zL'$ with the double alcove
$P_{3,p}$; if $\fwd_f\circ \zL'\subset P_{3,p}$ all $7$ points in
the fusion survive.  The truncations in \ft\ are precisely
described by (intersections of) orbits of the KW affine Weyl
group starting from a point on $\fwd_f\circ \zL' \cap\,  P_{3,p}$
and reaching points of $\fwd_f\circ \zL'$ outside of $P_{3,p}$.
Assigning (according to \ff\ ) multiplicity one to all points of
$\fwd_f$ (or to its isomorphic image $\fwd_f\circ \zL'$), the
points on any orbit contribute with a sign ${\rm det}(w)$
determined by the corresponding element $w=w_{\zb}$; here ${\rm
det}(w)$ is $1$ or $-1$ if the number of elementary reflections
of $W$ representing $w_{\zb}$ is even or odd.  To visualize these
orbits draw the big alcove for a given $p$ as in fig. 1. and
attach on it the shifted diagram $\fwd_f\circ \zL'$ -- the
shifted highest weight itself can be outside of the alcove.

Another example of this truncation mechanism is provided by the
fusions of $[1,1]$.
Taking a sufficiently large $\zL'$ in the fusion
$[1,1] \otimes \zL'$ -- i.e., $\zL'$ and $p$ such that
$\fwd_{[1,1]} \circ \zL' \subset P_{3,p}$, so that no truncations
occur, the FR multiplicities coincide with the weight
multiplicities of $[1,1]$ and can be computed from the general
formula \pvf.  
Alternatively the same weight multiplicities can be
recovered (and that is how originally we obtained the values
indicated on fig. 4), by  the above mechanism of truncation
along orbits of the KW group, given the FR multiplicities 
for ``smaller'' $\zL'$ and $p$.
\medskip

%%%%%%%%%%%%%%%%%%%%%%%%%%%%%%%%%%%%%%%%%%%%%%%%%%%%%%%%%%
%            section 6:
%%%%%%%%%%%%%%%%%%%%%%%%%%%%%%%%%%%%%%%%%%%%%%%%%%%%%%%%%%

\newsec{ ``Verma modules'' and  weight diagrams.
The role of KW affine Weyl group as a ``truncating'' group.}

We shall now formulate in a more concise general form the analog
of the FR multiplicities formula of \K, \MW, \FGP\ for integrable
representations ``inverting'' the argumentation of \FGP. As a
motivation for what follows recall that the $sl(3)$ finite
dimensional representations, equivalently their supports, i.e.,
the weight diagrams $\Gamma_\zl$, can be resolved via the action
of $\bW$ in terms of $sl(3)$ Verma modules.

First we introduce some more notation. Recall that $\ba_0=-\za_3\,,
\ba_i=\za_i\,, i=1,2\,.$  Given an affine root $\za$ among the
subset $\{\za_i\,, \za_{\hi}=\za_0 +\za_3-\za_i\,,$ $i=0,1,2\}$
there is a projection $\ba$ in the set of $sl(3)$ roots.
Accordingly $w_{\za}\in W$  projects to $\bw_{\za}=w_{\ba}\in
\bW$.  Vice versa given a  $sl(3)$ root and a corresponding
reflection there is an unique affine root among the above subset
and a corresponding reflection in $W$. Similarly for any set of
$sl(3)$  roots $w(\bS)\,, $ $w\in \bW\,,$
$\bS=\{\za_1,\za_2\}\,,$  there is a set $S_w$ comprised of a
pair of affine roots such that it projects to $w(\bS)$.  For even
(odd) length $w\in\bW$ the sets $S_w$ consist of roots in
$\{\za_i\}\,$ ($ \{\za_{\hi}\}\,$) $i=0,1,2$, respectively.
Let $Q^w=\oplus_{\za\in w(\bS)}\,\IZ_+\,\za$ for $w\in\bW$.  We
recover $Q^+$ for $w=id$.  For $w\in\bW$ of even length
let $T(w)$ be a group isomorphic to $\bW$ generated by the affine
reflections labelled by the roots in  $S_w\,:$
$T(id)\equiv\bW\,,$ $T(w_{21})$ is generated by $\{w_0,w_1\}\,,$
etc.  For odd length $w\in \bW$ the set $T(w)$ (which also
projects to $\bW$) is defined as $\{\un\,,w_{\hi}\,, w_{\hj}\,,
w_{j k}\,, w_{i k}\,, w_{k}\}\,,$ $\hat{k} \not =\hi,\hj \,, $ 
if $S_w=\{\za_{\hi}\,,\za_{\hj}\}\,, i,j=0,1,2\,,i\not =j\,.$  
E.g., $T(w_3)=\{\un, w_{\h2}, w_{\h1}, w_{10}, w_{20}, w_0\}\,.$

Recalling that the affine Weyl group $W$ is a semidirect product
of the horizontal (finite) Weyl group $\bW$ and the root lattice
$Q$, one can view the affine Weyl group graph $\wyg$ introduced
above as the translations of the basic hexagon, the graph of
$\bW$, by $\kappa Q$, i.e., action on the six vertices of the
basic hexagon by powers of $w_{i\hi}$ or $w_{\hi i}$, $i=0,1,2$.
More generally for $w\in\bW$ define
\eqn\mI{ 
\wyg_{\zL}^w = \cup_{w'\in T(w)}\, 
\Big(w'\cdot\zL + \kappa Q^{\overline{w'} \,w}\Big)\,, \qquad   w\in
\bW\,.  } 
In particular $\wyg^{id}_{\zL}$ coincides with $\wyg^+_{\zL}$
from above (see figs. 2,5),  while the subgraph
$\wyg^-_{\zL}:=\wyg^{w_3}_{\zL}$ is illustrated on fig. 6.

By analogy with the $sl(3)$ case where the set $\zl - Q^+$
describes the support of the Verma module of highest weight $\zl$
we shall refer to $\wyg^w_{\zL}\,$ as to ``Verma modules'' of
highest weight $\zL\,.$ The range of $\zL$ is not  confined to
the admissible domain \da\ and accordingly we can drop at this
point the requirement of rationality of the parameter $\kappa\,.$

In the case when the affine Verma module of highest weight
$\zL=-\zl \kappa$ or $\zL=w_3\cdot(-\zl \kappa)$ contains
by the Kac-Kazhdan theorem a singular vector of weight
$w_{\zb_i}\cdot\zL$ or $w_{\zb_\hi}\cdot\zL$, $i=1,2$ (i.e., $\zl$
is integral dominant, $\zl\in P_+\,,$ or strictly integral dominant, 
$\zl\in P_{++}\,$) then this weight belongs to
the graph $\wyg^{+}_\zL$ or $\wyg^{-}_\zL$ respectively. We can
view these weights as the highest weights of  ``Verma
submodules'' $\wyg^{w_i}_{w_{\zb_i}\cdot\zL}\,$ and $\wyg^{w_{i
3}}_{w_{\zb_{\hi}}\cdot\zL}\,,$ respectively.  One can extend
this to the  other four types of ''Verma modules'' with proper
infinite range of $\zL$ dropping the upper bounds of the alcoves.
Thus we  have six different analogs of the $sl(3)$ reducible
Verma modules (of integral dominant highest weights), the
singular vectors of which are governed by the corresponding KW
horizontal group $\bW^{\zl}$.

One can think of the ``Verma modules'' introduced
here as comprising some ``extended'' in the sense of \MFF\ states
given by compositions of noninteger powers of the three lowering
generators $E^{-\za_i}$, $i=0,1,2$, of $\hsl3$ -- these powers
are dictated by the (shifted) affine Weyl reflections indicated
on the graph $\wyg$, in particular any path from the origin to a
``singular vector'' corresponds to a MFF expression for the
``true'' singular vector of the affine algebra Verma module.
This intuitive picture is quite precise in the $\widehat{sl}(2)$ 
case where all states of the  corresponding ``extended'' Verma 
modules have multiplicity one. Then these modules are seen to 
be isomorphic to Verma modules of the superalgebra $osp(1|2)$; 
the two sublattices of the corresponding graphs,  associated with 
the two elements of the $sl(2)$ Weyl group, correspond to even or 
odd submodules with respect to $sl(2)$.

The problem of multiplicities however makes doubtful the
usefulness of such ``extended Verma modules'' interpretation in
our case and what replaces the above superalgebra is yet to be
seen. Instead we follow further the analogy with the $sl(3)$ case
and we  assign multiplicities $K_{\mu}^\zL$ to the weights $\mu$
of a general ``Verma  module'' of highest weight $\zL$, starting
from a multiplicity $1$ assigned to the points on the border
rays, and with each step inward increasing by $1$ the
multiplicity at the points of the subsequent pair of lines
parallel to them (going
through diagonals of the honeycomb hexagons) ;
see fig. 5,6 where this assignment is depicted. Not to overburden
the notation we shall not indicate explicitly the dependence of
$K_{\mu}^\zL$ on the type of module.  There are simple formulae
(to be presented elsewhere) for the generating functions of the
weight multiplicities assigned to any of the $6$ sublattices in
\mI\ which generalise the Kostant generating function for
$sl(3)$ Verma modules.

Let $\zL=-\zl \kappa\,, $ or $\zL=w_3\cdot(-\zl \kappa)\,,$ with
$\zl\in P_+\,,$ or  $\zl\in P_{++}\,,$ respectively.  We define
now ``weight diagram'' multiplicities $m_{\mu}^{\zL}$ according
to
\eqn\mII{
m_{\mu}^{\zL}=\sum_{w \in \bW^{
\zl}} \, {\rm det}( w)\, 
K^{w \cdot \zL}_\mu\,, \quad \mu \in \wyg_{\zL}^{\pm}\,,
}
and we shall refer to the (finite) collection of points $\{\mu\}$
with nonzero $m_{\mu}^{\zL}$ as ``weight diagram'' of $\zL$ to be
denoted $\G_{\zL}$. The proof that this definition has sense,
i.e., that the numbers in the l.h.s. of \mII\ are nonnegative
integers, extends the corresponding argument for the weight
diagrams of the finite dimensional representations of $sl(3)$ and
will appear elsewhere.  The previously discussed examples are
easily checked to fit the definition -- in fact  we have arrived
at the Verma module multiplicities $K^{\zL}_\mu\,$ ``inverting''
\mII, i.e., solving it for  small values of $K^{\zL}_\mu\,$
with the l.h.s. provided by the examples.  An example of a weight
diagram of the second kind (i.e., of highest weight
$\zL=w_3\cdot(-\zl \kappa)\,,$  $\zl\in P_{++}\,$)
is provided by the $4$-point diagram
of the representation $h=\,\[1,1\]\,$ -- the ``Verma modules
resolution'' formula \mII\ for this case is  illustrated in fig.
7, while figs. 2,3 illustrate \mII\ for $\zL=[1,0]$ and
$\zL=[2,0]$. On all these figures the positions of the singular
vectors (the two fundamental ones corresponding to
the simple reflections in $\bW^{\zl}$ and the ones corresponding to
compositions of such reflections) are indicated by $\star$. 

The two types of diagrams $\G_{\zL}\,$ are  illustrated furthermore
 on figs.
8,9. by two generic examples. View the vertices of $\G_\Lambda$
as lying on a set of ``concentric''  ``hexagons'' (which could
degenerate into triangles); they are drawn through diagonals of
the honeycomb elementary hexagons. The points on the outmost
hexagon have multiplicity 1, and with every step inward the
multiplicity increases by one if the layers are hexagons or
stays constant once the layers degenerate into triangles. This is
the direct generalisation of the rule for the $sl(3)$ weight
diagram $\Gamma_{\zl}$ multiplicities.  In general the
intersection of an weight diagram with any of the $6$ lattices in
\mI\ (for fixed $w=\un\,,$ or $w=w_3$)
decomposes into a collection of weight diagrams of $sl(3)$
irreducible representations.

The same pictures admit  another interpretation which may serve
as an alternative definition leading to \mII : refining by $3$
the lattice $Q \kappa$ (i.e., $p \rightarrow 3p$) and scaling by
$3$ the weights $\zl$ we can identify the two types of diagrams
of highest weight $\zL$ with standard $sl(3)$ weight diagrams
with some points removed.  Namely we map the highest weight
$\zL=-\zl \kappa\,$ ($\zl \in P_+$) to $i(\zL):=3 \zl\,,$ and 
$\zL=w_3\cdot(-\zl \kappa)\,$ ($\zl \in P_{++}$)  to 
$ i(w_3\cdot(-\zl \kappa))= 3 \zl -2 \rho\,,$ resp. The 
points removed from the $sl(3)$ weight diagrams originate from
the centers of the honeycomb  hexagons shaping $\G_{\zL}$, i.e.,
having at least $3$ ($2$) common links with $\G_{\zL}$, see figs.
8,9, where these points are indicated by squares.  
The weight multiplicities defined in \mII\ coincide with
the $sl(3)$ weight multiplicities of the corresponding surviving
points.

The same embedding applies to the supports of the Verma modules.
The  $sl(3)$ counterparts of the weights $\mu\in
\wyg^{\pm}_{\zL}$  are recovered from the highest weight $\zL$. 
More explicitly, the same definition of the map $i$ applies to
the points $\mu$  on the first, $w'=\un$ lattice in \mI,
$w=\un\,,$ containing the highest weight. The rest are recovered
from such points.  Namely for $\zL=-\zl \kappa$ and $\zm\in \zL +
Q^{+}\kappa\,, $  we have $ i(w\cdot
\zm)=i(\zm)+w^{-1}\cdot(0,0)\,,$ for $w\in \bW.$ The points
removed from the support   $3 \zl - Q^{+}$ of the  $sl(3)$ Verma
module of highest weight $3 \zl$ (not representing images of
$\wyg_{\zL}^{+}\,$) belong to the subset $\cup_{\alpha > 0} \,
\big(3 \zl - 3 Q^{+} - (3 - \la\alpha, \rho\ra)\, \alpha\big)\,.$
This rule can be also used to recover the images of the ``Verma
modules'' of highest weight $w_3\cdot(-\zl \kappa)$, representing
each point of ${\cal W}^-_{\zL}\,$ as $w\cdot \zm\,,$ with  $w\in
\bW$ and some $\zm\in -\zl \kappa + Q^+ \kappa\,.$ The definition
\mII\ of the weight diagram multiplicities and the  mapping into
$sl(3)$ representations extend to the other $4$ of the $6$
different types of ``Verma modules'' \mI,  taken with a proper
choice of the simple root system generating $\bW^{\zl}\,;$ we
omit here the details.  Apparently this map preserves (up to
permutation) the positions of the singular vectors, e.g.,
$i(w_{\zb_j}\cdot \zL)=w_j\cdot i(\zL)\,,$ $j=1,2\,,$
$i(w_{\zb_1} w_{\zb_2}\cdot
\zL)=w_{21}\cdot i(\zL)\,,$  etc., i.e. the action of the   KW
groups is ``converted'' under this map into  ordinary  Weyl group
action.  This together with the conservation of the
multiplicities $K_{\zm}^{\zL}=\bar{K}_{i(\zm)}^{i(\zL)}$, the
latter denoting the multiplicity of the weight $i(\zm)$ in the
$\sl$ Verma module of highest weight $i(\zL)$, converts \mII\
into the standard Verma modules resolution formula for  $\sl$
taken for a subset of weights.
 
Note that if $\zL$ is in addition an admissible highest weight at 
level $3/p - 3$ the above defined  weight $i(\zL)$ is an integrable
highest weight at level $3p-3$.

Because of the trivial triality of the $sl(3)$ counterparts
(their weights lying on the $sl(3)$ root lattice $Q$) all
diagrams of both types contain a ``middle'' point $\zL_0$ mapped
to the $sl(3)$ weight $(0,0)$. It is given by
$\Lambda_0=\sigma^{-\tau(\Lambda)}([0,0])\,,$ where we recall
that $\tau(\Lambda)=n_1-n_2$ mod $3$ for $\Lambda=[n_1,n_2]$ or
$\Lambda=\[n_1\,,n_2\]\,$; $\zL_0$  has a (maximal) multiplicity
$3$ min$(n_1,n_2)+1$ and $3$ min$(n_1,n_2)-1\,,$ respectively.

We have already discussed the ``shifted weight diagram''
$\G_{\zL} \circ \zL'$ assigned to a highest weight $\zL$ sitting
on the first leaf of $P_{3,p}$.  In general for $\zL\,, \,
\zL'\in P_{3,p}$ the shifted weight diagram $\G_{\zL} \circ \zL'$
is a diagram isomorphic to $\G_{\zL}\,,$ $\G_{\zL} \ni \mu=
w\cdot \zL \mapsto w\cdot (\zL\circ\zL') \in
\G_{\zL} \circ \zL'\,,$ $w\in W\,,$ with a reference point
(shifted highest weight) $\zL\circ\zL':=\Lambda+w(\Lambda')$ with
$w=\un$ $ (w=w_3)$ if $\Lambda$ is on the first (second) leaf of
$P_{3,p}$, respectively. The weight multiplicities of $\G_{\zL}
\circ \zL'$ are given by 
\eqn\mIII{
m^{\zL;\zL'}_{w\cdot (\zL\circ\zL')}:= m^{\zL}_{w\cdot \zL}\,,
\qquad w\in W\,,
}
where the r.h.s. is determined from \mII. Alternatively the
shifted weight diagram can be recovered from its middle point
$\sigma^{-\tau(\Lambda)}(\Lambda')\,$;
recall that the definition of the generating
element $\sigma(\Lambda')$ in \aut\ depends on the type of
$\Lambda'$.

The  mapping into $sl(3)$ diagrams preserving weight
multiplicities can be extended to the shifted diagrams
identifying in particular the $sl(3)$ shifted highest weight with
$i(\zL)+i(\zL')\,.$ Note that  if both $\zL\,,$  $ \,\zL'$ are
points on the $2^{\rm nd}$ leaf this weight originates from a
center of a (boundary) hexagon on  the shifted weight diagram
$\G_{\zL} \circ \zL'\,,$ i.e, from a point beyond $\G_{\zL}
\circ \zL'\,$ itself. The image of the shifted highest weight
coincides with $w(i(\zL))+i(\zL')\,$ (a point on the
$sl(3)$ shifted weight diagram), where $w=\un$ $ (w=w_3)$
if $\Lambda'$ is on the first (second) leaf, respectively.

We are now at a position to formulate the analog of the FR 
multiplicities formula in \K, \MW, \FGP.
\medskip

{\bf Proposition 3.} The FR multiplicities of (a triple of)
admissible weights on the double alcove $P_{3,p}$ are given by
the formula

\eqn\frs{
N_{\zL \zL'}^{\zL^{''}}=\sum_{w \in W^{\zl^{''}}} \, {\rm
det}( w)\, m^{\zL;\zL^{'}}_{w\cdot \zL^{''}}\,.
} 

At present this statement is rather a conjecture supported by a
lot of empirical data, in particular it is established if one of
the three highest weights in \frs\ coincides with the fundamental
weights $f$ or $f^*$. It has been also thoroughly checked for
small values of $p\ge 4$; \frs\ extends to the
 degenerated case $p=2\,,\, p'=3$ excluded so far.

\medskip

An example illustrating \frs\ with $\zL$ sitting on the second
leaf of the alcove is provided by the representation
$h=\[1,1\]\,.$  Its weight diagram appeared in fig. 7. and its
fusions read\foot{This simplest representative of the $2^{\rm
nd}$ leaf admissible representations has to be included along
with $\{f\,, f^*\}$ in the fundamental set generating the fusion
ring.}
\eqn\hf{
h \otimes \zL' = 2\, \zL' \oplus w_1\cdot \zL'\oplus w_2\cdot
\zL' 
\oplus w_0\cdot \zL'
}
for any representations $\zL'$ on the second leaf and for generic
(i.e., not on the border lines of the alcove) representations
$\zL'$ on the first leaf. For  border representations on
the first leaf \hf\ reduces to three multiplicity one terms (or
to one such term if $\zL'=\sigma^l(\un)$).  Apparently the latter
truncation is a result of a $2$-points orbit leading to
$m_{\zL'}-m_{w_i \cdot \zL'}=2-1=1$ for $\zL'$ such that $w_i
\cdot \zL'\not \in P_{3,p}\,,$ for some (one) $ i=0,1,2\,$  or to
a $3$-points orbit for $\zL'=\sigma^l(\un)\,,$ in which case $w_i
\cdot \zL'\not \in P_{3,p}\,,$ for at least two $ i=0,1,2\,$. 
The example $\zL'=\sigma^2(\un)$ is illustrated on fig. 10.

For another example see fig. 11 and the Appendix.  Note that
unlike the integrable case there are no analogs here of
``$q$-dim$=0$'' representations since the walls of the 
admissible alcove and its images by the action
of the KW group  do not support (being on the lattice dual to 
the graph  ${\cal W}$) weights of ``classical'' representations.
\medskip

The nonnegativity of the FR multiplicities in \frs\ is ensured 
in general  by the fact that the r.h.s of \frs\ can be expressed in
terms of $sl(3)$ weight diagram multiplicities and the ordinary
affine Weyl group $W$ adapting the map $i$ discussed above.  One
recovers in this way the r.h.s. of the formula in \K, \MW,
\FGP\ for the FR multiplicities $N_{i(\zL)
i(\zL')}^{i(\zL'')}(3p)$ for particular triples of triality zero
highest weights in an integrable theory at level $3p-3$. 
(Recall that the Verlinde multiplicities are $\IZ_3$ graded.) The
details will be presented elsewhere, here we only illustrate this
statement with an example depicted on fig. 12, see the Appendix
for more details. The essential property used is that the
(shifted action) Weyl group orbit of a ``removed'' point contains
only ``removed'' points.  This reformulation of \frs\ is a
generalisation of the same type of relation between admissible
and integrable representations FR multiplicities in the case of
$\widehat{sl}(2)$ pointed out in (4.1) of \FGPa; in particular we
expect that the general fusion rules for $p'\ge 3$ will admit a
similar factorised form in terms of fusion
multiplicities for integrable  representations at levels $p'-3$
and $3p-3\,,$ see Appendix.

\medskip

The formula \frs\ admits as in \FGP\ a  reformulation in terms of
the analogs $m_{\zL \zL'}^{\zL^{''}}$ of the $sl(3)$ tensor
product multiplicities. It applies to weights
$\zL=w\cdot(-\zl \kappa)\,,$ $w\in \bW\,,$ $\zl\in P_+^{w}\,,$ where
$P_+^{\un}=P_+\,,\, P_+^{w_3}=P_{++}\,,$ and 
$P_+^{w_i}\equiv P_+^{w_{i3}}
=\{\zl\in P_+\,, \, \la \zl, \za_i\ra \not =0\}\,,$
$i=1,2\,.$ In particular for $\zL, \zL'$ of the type 
$w\cdot(-\zl \kappa)\,, w=\un\,, \,w_3\,,$
the multiplicities $m_{\zL \zL'}^{\zL^{''}}$  are defined 
as  in \frs, with  the affine KW group
$W^{\zl^{''}}$ replaced by its horizontal counterpart
$\bW^{\zl^{''}}\,,$ generated by $w_{\zg_i}\,,$ 
$\zg_i:=\la\zl^{''},\za_i \ra
\zd + w(\za_i)\,,$ $i=1,2\,,$ for $\zL=w\cdot(-\zl^{''} \kappa)\,,$ 
$w\in \bW\,.$
It has been checked on numerous examples that
the latter definition of $m_{\zL \zL'}^{\zL^{''}}$ makes sense,
in particular leading to the conservation of the ``classical''
dimensions computed from the weight diagrams.  We conclude with
some of the simplest examples
\eqn\Va{\eqalign{
f\otimes f^* &= \un \oplus  h \oplus [1,1]\,,
\ \ (7\times 7=1+5+43)\,,\cr
f\otimes f& = f^* \oplus  [2,0] 
\oplus w_1\cdot
[2,0]\,, \ \ (7\times 7=7+19+23)\,,\cr
h\otimes h &=\un \oplus 2\, h 
\oplus w_{1}\cdot h
\oplus w_{2}\cdot h
 \,,\ \ (5\times 5=1+2.5+7+7)\,,\cr
h\otimes f&=f\oplus w_{1}\cdot f
\oplus \[2,1\] 
 \,,\ \ (5\times 7=7+5+23)\,.
}}
This is a strong indication that our weight diagrams are the
supports of finite dimensional representations of a ``hidden''
algebra. Its $q$-version at roots of unity would provide
eventually a truncated tensor product equivalent to the fusion
rules. 
\medskip

\medskip

\newsec{Summary and conclusions.}

There are several different descriptions and derivations of the
fusion rules at integer level. No one of them is easily
extendable to the case of fractional level. Instead we have used
a combination of several methods -- neither of them is completely
rigorous or thorough at present.

We started with the method which directly generalises the
approach exploited  by \AY\  in the derivation of the admissible
representations fusion rules in the $\widehat{sl}(2)$ case. This
is the singular vectors decoupling method applied to $3$-point
functions which  works for fusions resulting in trivial (zero or
one) fusion multiplicities.  In particular it enabled us to
determine the fusion of the (fractional) fundamental
representations $f=[1,0]$ and $f^*=[0,1]$ with an arbitrary
generic admissible representation $\zL'$. The notions of weight
diagram and shifted weight diagram described by a set of words in
the affine Weyl group naturally arise as generalisations of the
$sl(3)$ counterparts.

Unlike the $sl(2)$ case the decoupling method is technically
rather involved due to the complexity of the general MFF singular
vectors and  presumably it has to be further elaborated in order
to treat cases with nontrivial multiplicities.  Instead we have
followed a strategy influenced to some extent by the study \Pas,
\DFZ\ of  graphs (generalising the ADE Dynkin diagrams) related
to  nondiagonal modular invariants of the integrable WZNW
theories.  Namely selecting and analysing a subset of the
decoupling systems of equations, corresponding to representations
on the  border paths of the  admissible alcove, we have
determined (under some additional assumptions) for $p'=3$ and any
$p\ge 4$ a fundamental fusion graph described by the fusion
matrix $N_f$. This allowed us to write a formula, borrowed from
\Pas, \DFZ,  for the general FR multiplicities at level $3/p -
3\,.$ It generalises the Verlinde formula in which the symmetric
modular matrix is replaced by the eigenvectors matrix
$\psi_a^{(\mu)}$ of $N_f$.  It is highly nontrivial that a
formula of this kind gives nonnegative integers -- we have no
general proof of this fact checked for small values of $p$.

Knowing the  matrix  $N_f$ one can determine in principle
for any $p$ its eigenvectors -- yet the proposed Pasquier  --
Verlinde type formula \pvf\ is still not very explicit in view of
the absence so far of a general analytic formula for
$\psi_a^{(\mu)}$. So our next step was to look for an alternative
formula for FR multiplicities, generalising our old work in \FGP,
see also \K, \MW.  While the previous  approach can be looked at as
related to a resolution of the irreducible admissible
representations in terms of a kind of generalised reducible
``Fock modules'' (since the differential operators realisation of
the generators of $\hsl3$ is equivalent to a generalised free
field bosonic realisation) the formula \frs\ described in section
6 rather relies on the idea of ``Verma modules'' resolution.

We recall that the starting point in \FGP\ was the standard Weyl
formula for the multiplicities of irreps in tensor products of
$sl(n)$ finite dimensional representations. This formula involves
the weight multiplicities of $sl(n)$ finite dimensional
representations, i.e., their weight diagrams, which can be
recovered by resolution of $sl(n)$ Verma modules. So it was
natural to try to interpret similarly the generalised weight
diagrams  we have encountered in section 4 -- formula \mII\ is
precisely of this type with the Weyl group $\bW$ replaced by the
horizontal KW group $\bW^{\zl}$.

The hard part in this alternative approach is the absence
initially of an obvious candidate for the finite dimensional
algebra whose representation theory matches the structures
introduced in section 6.  Instead we have described the ``Verma
modules'' by their supports, i.e., the set of weights and their
multiplicities. Our final step was as in \FGP\ to ``deform''  the
classical formulae    replacing the horizontal Weyl group
with its affine analog, in our case the affine KW group
$W^{\zl}$.  While in \FGP\ we have been guided at this step by
the representation theory of the deformed algebra $U_q(sl(n))\,$
for $q$ -- a root of unity, here once again we lack so far the
$q-$ counterpart of the hidden algebra -- rather the consistency
of \frs\ with the alternative approaches suggests the existence
of such  deformation.

The emerging  finite dimensional (graded)  algebra (and its $q$
-- counterpart) behind the series $\kappa=3/p\,$ of $A_2^{(1)}$
admissible representations is the most interesting outcome of
this work.  It might be possible to recover this algebra
(containing $sl(3)$ as a subalgebra) from the supports of the
(reducible)  Verma modules introduced above.  In fact there is
some evidence that the algebra is encoded by the $43$ -
dimensional representation $[1,1]$, the fractional "adjoint"
representation. In particular the map $i$ discussed in section 6
provides a natural way of introducing a root system related to
the weight diagram of this representation. While a subset of
these roots sits on the $sl(3)$ weight lattice $P$ (which
contains the $sl(3)$ root lattice $Q$), there are ``fractional''
roots beyond it.

Another remaining related problem is the description of the
characters of the finite dimensional representations of this
algebra and their $q$-version, i.e., the derivation of an
explicit formula for the eigenvectors  matrix $\psi_a^{(\mu)}$.
This would give more substance to the analog
\pvf\ of the Verlinde formula  and would eventually allow to
prove its equivalence to \frs, so far checked only on examples.

Starting from an explicit finite dimensional algebra may simplify
also a more abstract derivation of the $\widehat{sl}(3)$
admissible representations fusion rules as well as their
generalisation to $\widehat{sl}(n)$.

These questions are under investigation.

Let us finally mention that the problem of deriving the
admissible  representations fusion rules  might be relevant for
the analogous problem for representations of $W$-algebras
obtained by quantum (non principal) Drinfeld--Sokolov reduction
from the affine algebras -- see  the analogous reduction of Verma
modules singular vectors in \FGPc.

\bigskip

{\bf Acknowledgements}
\medskip
We would like to thank V. Dobrev, V. Molotkov, Tch. Palev, I.
Penkov,  V. Schomerus and V. Tolstoy  for useful discussions.
A.Ch.G. and V.B.P. acknowledge the hospitality and the support of
INFN, Trieste.  A.Ch.G. acknowledges the support of the Alexander
von Humboldt Foundation.  This paper was partially supported by
the Bulgarian Foundation for Fundamental Research under contract
$\Phi-404-94$.

\medskip

\appendix{A}{Examples}

 On  {figs.} 11 we  illustrate the fusion
$$
[[2,1]]\otimes [0,1] = [[2,2]]\oplus [1,1] \oplus [[1,a]] \oplus
[[1,1]] \oplus [a,1]\,, \quad a=p-2 \,.
$$
\noindent
On the first fig. 11a we depict the weight  diagram ${\cal
G}_{\Lambda}$ for $\Lambda=[[2,1]]$; the multiplicities are
indicated in circles.  On fig. 11b we depict the same diagram but
shifted by $\Lambda'=[0,1]$, i.e., the shifted diagram ${\cal
G}_{\Lambda}\circ\Lambda'$ having $\sigma^2(\Lambda')=[1,a]$ at
the middle point; for short  $a=p-2$, $b=p-3$.  The shifted
diagram is situated on the graph ${\cal W}$ viewed as a weight
lattice. All weights sitting outside the addmissible alcove are
indicated by (??) and the weights of the first and the second
leaf are indicated by circles and boxes, respectively. Reflecting
in the dotted lines (the walls of the alcove) gives the
truncations.  The weights [[a,1]] and [1,a] get their
multiplicities truncated to $0=2-2$; [b,0] to $0=1-1$; [0,0] to
$0=2-1-1$; [[1,1]] and [1,1] to $1=2-1$.

\medskip
\medskip

On fig. 12 we illustrate the relation between the fusion rules of
admissible representations at level $3/p -3$ and of the
integrable representations at level $3p-3$.  We compare the
fusions  $[1,1]\otimes[1,0]$ and $(3,3)\otimes(3,0)$ choosing $p=4$.

The shifted weight diagram ${\cal G}_{[1,1]}\circ [1,0]$ is
mapped into the weight diagram $\Gamma_{(3,3)} +(3,0) +\rho$, the
shifted highest weight $\mu=[1,1]+[1,0]=[2,1]$ is identified on
the picture with the $sl(3)$  shifted highest weight $(6,3)
+\rho=(7,4)$.  The reason we shift here in addition  the
integrable weight with $\rho$ is that following the tradition we
describe the fusion of integrable representations geometrically
by ordinary, nonshifted Weyl group action, but acting on the
shifted by $\rho$ weights; in the final result all surviving
weights are shifted back with $-\rho$. The boundary triangle
enclosing the admissible alcove  $P_{3,4}$ is mapped into the
boundary lines of the $\hsl3$ integrable alcove $P_{+}^{12}$.
The origin $[0,0]$ of $P_{3,4}$ corresponds to the $sl(3)$ weight
$(10,1)\,,$ i.e., the corner point $\sigma((1,1))$ of the alcove
$P_{++}^{12}$.  (For $p=5$ it would have corresponded to $(1,13)=
\sigma^{*}((1,1))\,$  in $P_{++}^{15}$.)

The ''removed'' weights sitting at the centers of the elementary
hexagons are indicated by squares. Such are in particular the
weights $\lambda''$ , $\lambda''+\rho = (7,4)-2\alpha_1\,,
(7,4)-2\alpha_3 -\alpha_1\,, (7,4)-\alpha_3\,$ appearing with
multiplicity one in the product $(3,3) \otimes (3,0)\,.$ The
remaining $7$ weights $\zl''$ in this product, $\zl''+\rho=
(7,4)\,, (5,5)\,, (1,7)\,, (4,4)\,, (5,2)\,,(4,1)\,, (2,5)\,,$ 
have admissible counterparts all of
multiplicity one,  recovered by the inverse map $i^{-1}$.  E.g.,
$i([2,1])=(6,3)\,,$ and hence $i^{-1}((6,3))=[2,1]\,.$ Similarly
$i^{-1}((3,0))=[1,0]=w_{0 \h0}\cdot [2,1]\,,$ $i^{-1}((0,6))=[0,2]=
w_{\h1 1}\cdot [2,1]\,.$ Furthermore $
i(w_1\cdot [2,1])= i([2,1])-\alpha_1=(6,3) -\za_1=(4,4)\,,$ hence
$i^{-1}((4,4)) = w_1\cdot [2,1]= \[1,2\]$.  Similarly
$i^{-1}((4,1)) = i^{-1}((6,3) - 2\za_3) =w_3\cdot
[2,1]=\[2,1\]\,,$ and $i^{-1}((1,4))=w_2\cdot [0,2]=w_{021}\cdot
[2,1]=\[2,2\]\,,$ 
$\,i^{-1}((3,3))=i^{-1}((6,3)-\za_1-\za_3)=w_{21}\cdot 
[2,1]=[1,1]\,.$

The final result can be checked to coincide with what is prescribed
by \ff.  It can be cast into the form
\eqn\aI{
\zL_1 \otimes \zL_2 = {\sum_{\zl_3}}^{'}\, 
 N_{_{3\zl_1\, 3\zl_2}}^{^{\zl_3}}(3p)\, 
\, \ w^{(\zl_3)}\cdot(\zL_1+ \zL_2)\,,
}
where $w^{(\zl_3)}\in W\,$ is determined from
$i\big(w^{(\zl_3)}\cdot(\zL_1+ \zL_2)\big)=\zl_3$ as described
above, the multiplicity in the r.h.s. is the integrable FR
multiplicity at level $3p-3$ and the prime in the sum indicates
that all "removed" points are excluded.

By analogy with (4.1) of \FGPc\ the expected generalisation for
arbitrary $p'$ and admissible ($1^{\rm st}$ leaf) highest weights
$\zL_i=\zl_i^{'}-\zl_i
\kappa\,, i=1,2$ reads
\eqn\aII{
\zL_1 \otimes \zL_2 = \sum_{\zl_3'}\,{\sum_{\zl_3}}^{'}\,
N_{\zl_1', \zl_2'}^{\zl_3'}(p')\, 
 N_{_{3\zl_1\, 3\zl_2}}^{^{\zl_3}}(3p)\, 
\, \ w^{(\zl_3)}\cdot(\zl_3'-\kappa(\zl_1+ \zl_2))\,,
}
where $N_{\zl_1', \zl_2'}^{\zl_3'}(p')\, $ is 
the integrable FR multiplicity at level $p'-3$.

\bigskip

\listrefs

\bye